\documentstyle[12pt]{article}

\begin{document}

  \baselineskip=18.6pt plus 0.2pt minus 0.1pt


  \makeatletter
  \@addtoreset{equation}{section}
  \renewcommand{\theequation}{\thesection.\arabic{equation}}
  \begin{titlepage}
  \title{
  \hfill\parbox{4cm} {\normalsize UFR-HEP/02-09}\\ \vspace{1cm}
       {\bf     F-theory Duals of M-theory on   $G_2$ Manifolds
            from Mirror Symmetry }
  }
  \author{
  Adil Belhaj \thanks{{\tt ufrhep@fsr.ac.ma}}
    {}
  \\[7pt]
  {\it  High Energy Physics Laboratory,  Physics Department, Faculty
  of sciences}\\{ Avenue Ibn Battouta, PO Box 1014, Rabat, Morocco}
  }

  \maketitle \thispagestyle{empty}
  \begin{abstract}
   Using   mirror  pairs $ ( M_
   3, W_3)$ in type II superstring
    compactifications  on Calabi-Yau threefolds, we
    study, geometrically,
   F-theory duals of M-theory on  seven manifolds with $G_2$ holonomy.  We first
  develop
   a  way for getting  Landau Ginzburg (LG) Calabi-Yau   threefolds $W_3$,
    embedded in four complex dimensional
    toric varieties,
     mirror to  sigma model on toric Calabi-Yau threefolds $M_3$.
     This method gives directly the right dimension without
     introducing non dynamical variables.
     Then,
     using toric  geometry tools,   we discuss     the  duality
     between M-theory on ${S^1 \times M_3}\over {\bf Z_2}$  with  $G_2$ holonomy
      and F-theory  on  elliptically  fibered
    Calabi-Yau fourfolds with $SU(4)$ holonomy, containing   $W_3$ mirror
  manifolds.
    Illustrating examples are presented.
     \end{abstract}
  \newpage
  \newpage
  \end{titlepage}
  \newpage
  \def\be{\begin{equation}}
  \def\ee{\end{equation}}
  \def\bea{\begin{eqnarray}}
  \def\eea{\end{eqnarray}}
  \def\nn{\nonumber}
  \def\l{\lambda}
  \def\t{\times}
  \def\[{\bigl[}
  \def\]{\bigr]}
  \def\({\bigl(}
  \def\){\bigr)}
  \def\p{\partial}
  \def\o{\over}
  \def\ta{\tau}
  \def\cm{\cal M}
  \def\R{\bf R}
  \def\b{\beta}
  \def\a{\alpha}
  \newpage
  \section{Introduction}
  Over the few past years, there has been an increasing interest in
  studying string
    dualities. This interest is  due to the fact that this subject  allows
    to explore several connection between different models in string
    theory. Interesting  examples are  mirror symmetry between pairs of Calabi-Yau
  manifolds
    in type II superstrings \cite {1,2,3,4}
    and  strong/weak coupling duality among this type and heterotic superstrings
  on
    $K3\t
    T^2$ \cite{5}.
       The  most important consequence of  the study  of  string duality  is that
  all
   superstring models are equivalent in the sense that they
   correspond to different limits in the  moduli space of the same
   theory, called M-theory  \cite{6,7,8,9}. The latter,  which is considered
  nowadays
    as the best candidate
   for the unification of the weak and strong coupling sectors of
    superstring models,
     is described,
    at low energies, by an eleven dimensional
   supergravity theory. \par
    More recently,   a special interest has been given to  the study  of the
  compactification of
       the M-theory on seven real manifolds $X_7$ with non trivial
    holonomy  providing  a potential point of contact with low energy semi
  realistic
    physics in our  world \cite{9,10}. In particular, one can obtain four
    dimensional theory with $N=1$ supersymmetry by compactifying
    M-theory  on $R^{1,3}\t X_7$ where $X_7$ a seven manifold
    with   $ G_2$  exceptional holonomy [10-23]. This  result can be understood from   the
    fact that the $ G_2$ group is the maximal subgroup of $ SO(7)$
    for which  a eight dimensional spinor of it can be  decomposed  as a
    fundamental of $G_2$ and one  singlet.
  In this regards, the  $N=1$ four dimensional resulting physics
  depends on geometric properties of  $X_7$. For instance, if
   $X_7$ is smooth, the low energy theory contains, in addition
    to
   $N=1$
   supergravity, only abelian gauge group and neutral chiral
   multiplets. In particular, one has $b_2(X_7)$ abelian vector multiplets
   and $b_3(X_7)$ massless neutral chiral multiplets, $b_i$ denote
   the Betti numbers of $X_7$.  The non abelian  gauge symmetries with chiral
  fermions   can be
  obtained
    by considering limits where $X_7$ develops singularities
    \cite{15,16,17}. The  $N=1$ theory in four dimensions   can be also obtained  using alternative  ways.  A way is to consider
    the $E_8 \t E_8$ heterotic  superstring
      compactifications.   In this way, the compact manifold is a Calabi-Yau
  threefold with
      SU(3) holonomy  with an appropriate  choice of vector bundle over it braking
  the $E_8 \t
      E_8$ gauge symmetry \cite{20}.\\
      Another way, which is  dual to the  heterotic compactification, is to
  compactify
       F-theory on elliptically fibered  Calabi-Yau fourfolds with $SU(4)$ 
  holonomy group [25,26,27,28].
       At this level, one might naturally ask the following questions. Is
  there a duality between  M-theory  on seven $G_2$ manifolds and
  F-theory on Calabi-Yau fourfolds and what will be the geometries
  behind this duality? \\ In this paper,  we address these questions
  using  toric geometry and mirror pairs in type II superstrings
  propagating  on Calabi-Yau threefolds.   In particular, we discuss
  the duality between M-theory on $G_2$ manifolds  and F-theory on
   elliptically fibred Calabi-Yau fourfolds with $SU(4)$ holonomy.
   In this study, we  consider M-theory on spaces of the form $ {S^1 \times
   M_3}\over {\bf Z_2}$  with $G_2$ holonomy where $M_3$ is  a   local  Calabi-Yau
   threefolds
   described physically by $ N=2$ sigma model in two dimensions.  The F-theory
  duals of
   such models can be obtained by the compactification
   on $ {T^2 \times
   W_3}\over {\bf Z_2}$, where $ (M_3, W_3)$ are mirror pairs in type
   II superstring compactifications.   More precisely, using toric geometry tools,
  we first
     develop   a way for getting LG  Calabi-Yau
   threefolds $W_3$ mirror to sigma model on toric Calabi-Yau $
   M_3$. This method is based on solving the mirror constraint eqs
   for LG theories in terms of the toric data of sigma model on $
   M_3$. Then  we give    a toric
   description of  the above mentioned duality.  In particular, we
    propose a special ${\bf Z_2}$  symmetry acting on the toric
   geometry angular  variables producing  seven manifolds with K3
   fibrations  in  $G_2$ manifolds compactifications.  \\
   The organization of this  paper is as follows.  In section 2, 
    using
    mirror symmetry in type II superstrings, we
    propose a possible duality between M-theory on $G_2$ manifolds
    and F-theory on  elliptically fibered  Calabi-Yau fourfolds
    involving  mirror pairs  of Calabi-Yau thereefolds $(M_3, W_3)$.
    In section 3 we  develop  a method
    for getting LG  Calabi-Yau
   threefolds $W_3$ mirror to sigma model on toric Calabi-Yau $
   M_3$. This method is based on solving the mirror constraint eqs
   for LG Calabi-Yau  theories in terms of the toric data of sigma model on $
   M_3$,  giving  directly the right dimension of
   the mirror geometry without introducing  non dynamical variables.
   In this way,  the mirror LG
   Calabi-Yau threefolds $W_3$
   can be described as
    hypersurfaces in four dimensional weighted projective spaces  ${\bf WP^4}$,
  depending
     on the toric data of $M_3$. In section 4 we  give  a toric
   description of  the above mentioned duality. In particular, we
    propose a special ${\bf Z_2}$  symmetry acting on the toric
   geometry  coordinates leading to   $G_2$ manifolds  with K3
   fibrations. Then
    we  discuss others examples where  ${\bf Z_2}$ acts trivially on  mirror
    pairs $ (M_3, W_3)$. In the last section,   we give our
    conclusion.
  \section{ On F-theory duals of  M-theory on $G_2$-manifolds in four
  dimensions}
   In this section we  want to  study  the duality between  M-theory on
  $G_2$-manifolds
   and F-theory on elliptically   fibered  Calabi-Yau fourfolds. To start,  recall
  that
    the duality between M-theory and F-theory  was studied
   using  different ways. For instance, this can be achieved using  Mayr work
  based on
   the local  mirror symmetry and special  limits in the elliptic compactification
  of  F-theory on 
    Calabi-Yau manifolds [29,30].
    Alternative approach  can be  done using the   Ho$\check{r}$ava$-$Witten
  compatification
    on spaces of the form
  ${S^1 \over Z_2}\times Y$, where $Y$ is a Calabi-Yau threefolds,
  giving  rise  to $N=1$ supersymmetry in four dimensions \cite{27}.
  The latter involves  a weak coupling limit given by the heterotic
  superstring compactified  on  Calabi-Yau threefolds which may have
  F-theory dual on Calabi-Yau  fourfolds.  However, in this  work we
  want to introduce  manifolds with  $ G_2$ holonomy in  the game.
  In particular we would like to discuss  a new  duality between
  M-theory on $G_2$-manifolds  and F-theory on elliptically fibred
  Calabi-Yau fourfolds, with SU(4) holonomy group in four dimensions
  with $ N=1$  supersymmetry. Before going ahead, let us start by
  the first possible equivalence between M-theory and F-theory. This
  can appear in  nine dimensions by help of T-duality in type II
  superstrings. Roughly speaking, this duality can be rewritten,
  using the above mentioned theories, as follows
  \be
   \mbox{M-theory on}\; S^1 \t S^1({R})= \mbox{F-theory  on}\; T^2\t S^1({1 \over
   R}),
  \ee where $ R$ and ${1 \over R}$ are the raduis of type II one
  circle compactifications. In this case, the $Sl(2,{\bf Z})$
  symmetry in type IIB superstring can also have a geometric
  realization in terms of the M-theory on elliptic  curve $ T^2$. Duality (2.1) can be
  pushed  further for describing the same phenomenon involving
  spaces that are more complicated than a circle, such as Calabi-Yau
  spaces in which the T-duality will be replaced by the  mirror
  transformation. Indeed,  using mirror symmetry duality in the
  Calabi-Yau compatifications, eq(2.1) can be extended to
  \be
   \mbox{M-theory on}\; S^1 \t M_3= \mbox{F-theory  on}\; T^2\t W_3,
  \ee where $(M_3,W_3)$ are
                mirror pairs  manifolds whose Hodge numbers
  $h^{1,1}$ and $h^{2,1}$  satisfy
     \bea
  h^{1,1}(M_3)= h^{2,1}(W_3)\nn\\ h^{2,1}(M_3)=h^{1,1}(W_3)\nn.
   \eea
    In this way,  the complex (Kahler) structure  moduli space of  $M_3$ is
            identical to the Kahler  (complex) structure  moduli space of
            $W_3$ and  the above four dimensional
  models  are  equivalent to  type II superstrings compactfied on
  mirror pairs $(M_3, W_3)$.  Thus, 
    equation (2.2)  describes  models  with  eight supercharges in four
  dimensions, ie $ N=2$ 4D.   In what follows, we want to relate
  this duality to M-theory on manifolds with $G_2$ holonomy.
  However,  to do this one has to break the half of the
  supersymmetry and should look for the expected holonomy groups
  which needed in both sides. It turns out that  there are some
  possibilities to realize the first requirement. One way is to use
  the result of the string compactifications on Calabi-Yau manifolds
  with
   Ramond-Ramond fluxes  \cite{28}. Another method of doing,  we are interested in 
  here, is
  to  consider the modding of the above duality  by  $\bf Z_2$
  symmetry.  Note, in passing, that this operation  has been used in
  many cases in string theory compatifications to break the half of
  the supercharges, in particular,  in the case of  type IIA
  propagating on $T^4$, which known by an orbifold limit of $K3$
  surfaces.  The four dimensional M-theory/F-theory dual pairs with
  $ N=1$ supersymmetry can be obtained by $\bf Z_2$ modding of
  corresponding dual pairs with $ N=2$ given in (2.2). Using this
  procedure, the resulting space in M-theory compactification  is
  now a quotient space of the following form
  \be
  X_7 ={S^1\times M_3 \over {\bf Z_2}},  \ee
    where  ${\bf Z_2}$ acts on $S^1$ as a reflection and   non
  trivially on the Calabi-Yau threefolds $M_3$.   The holonomy group
  of this geometry is  now larger than $SU(3)$ holonomy of $ M_3$,
  which is the maximal subgroup of $G_2$  Lie group.  The  
  superstrings propagating on this type of manifolds preserve 1/8 supercharges in three dimensions. Using the
  decompactification mechanism, M-theory on this  geometry has
  similar feature of seven manifolds with $G_2$ holonomy. This type
  of manifolds  has  been a subject to an intensive interest
  during the last few years dealing with different problems in superstring theory. In particular, these  involve, the computation of instantons superpotentials \cite{ 101},    the description  of  IIA  superstring  orientifold  compactifications giving four dimesional $N=1$  models \cite{102}  and  the study   two dimesional  superconformal field theories \cite{18}. In all  those  works, the $\bf Z_2$  acts   on the complex homogeneous  variables,   defining the Calabi-Yau threefolds,  by complex conjugation.  However,   in this present  work  we will use  a new transformation  acting on the toric geometry realizations of Calabi-Yau spaces. In  section four,  we will show that this procedure   leads  to K3 fibrations in G2 manifold  compactifications. In this way, we  will  be  able to find  heterotic superstring  dual models using  M-theory/heterotic duality in seven dimensions [ 10,17,18].\\ On F-theory side, the $
  N=1$ dual model  may be obtained using the same procedure  by
  taking  the following quotient space
   \be
   \quad W_4 ={T^2\times W_3 \over
  {\bf Z_2}}. \ee  In this equation,  ${\bf Z_2}$  acts nontrivially on the
  mirror geometry $ W_3$ and as a reflection  on $ T^2$ as follows
  \be
  {\bf Z_2}: \quad dz  \to -dz \ee where $z$ is the complex
  coordinate of the elliptic curve  $T^2$.  If this symmetry has some fixed points, they
  need to be deformed for obtaining a smooth manifold. This manifold
  is then elliptically  fibered over  $ W_3 \over {\bf Z_2}$. Like
  in M-theory side, the holonomy group of this  quotient space is
  larger than $SU(3)$ holonomy of $ W_3$. However, the four
  dimensional field theories with $N=1$ supersymmetry obtained from
  F-theory compactification require that  $W_4$ is an elliptically
  fibred Calabi-Yau fourfolds with $K3$  fibration with $SU(4)$
  holonomy. From these physical  and mathematical  arguments, we can
  propose, up some details, the following new  duality
  \be
   \mbox{M-theory on}\; X_7(G_2)= \mbox{F-theory  on}\;  W_4(SU(4)),
  \ee where in both sides involve mirror  pairs of Calabi-Yau
  threefolds $(M_3, W_3)$. In what follows, we want to give  a
  comment concerning  the relation connecting the   Betti numbers of
  the above  quotient  manifolds. This can be done just  by knowing
  the results of the numbers of vector multiplets  and massless
  neutral chiral multiplets obtained  from both side.  Using the
  results of \cite{29,30},  we expect the following formula \bea
  b_2(X_7)&= &h^{1,1}(W_4)-h^{1,1}(B_3)-1+h^{2,1}(B_3)\nn\\
  b_3(X_7)&=& h^{1,1}(B_3)-1+h^{2,1}(W_4)-h^{2,1}(B_3)+h^{3,1}(W_4),
  \eea where $B_3= {W_3 \over {\bf Z_2}}$.   These
   Betti numbers, in general,  depend   on  the
   framework  of  the
  geometric construction of  manifolds and  how  the ${\bf Z_2}$
  symmetry acts on the  Calabi-Yau threefold.  In this  present
  context, the no zero Betti numbers correspond to  the ${\bf Z_2}$
  invariant forms of the above quotient spaces. In next section, we
  will use the toric geometry language, its relation to sigma model
  and LG mirror geometries to give expected relations and give some
  illustrating examples. In this analysis, one has the following
  Hodge constraint equation \be h^{2,1}(M_3)=h^{1,1}(W_3)=0
   \ee
  reducing   (2.7)  to \bea b_2(X_7)&=
  &h^{1,1}(W_4)-1+h^{2,1}(B_3)\nn\\
  b_3(X_7)&=&h^{2,1}(W_4)-1-h^{2,1}(B_3)+h^{3,1}(W_4). \eea More
  precisely, our strategy will be as follows: \\ (i) Building of
  three dimensional Calabi-Yau threefolds and theirs mirror in terms
  of hypersurfaces in four dimensional toric varieties using the
  toric geometry technics, its relation to sigma model and LG
  theories. In this steep, we give  a tricky method for getting the
  mirror toric Calabi-Yau threefolds $W_3$, involved in  F-theory
  compactifications, from the toric data of $M_3$ manifolds.\\ (ii)
  We give a toric  description for the duality given in (2.6).
  \section{ Toric geometry  for Calabi-Yau  threefolds and theirs mirrors}
  \subsection { Toric geometry of Calabi-Yau threefolds $M_3$}
    Complex Calabi-Yau varieties  are Ricci flat spaces
     with a vanishing first Chern class $c_1=0$. They are an  important
     ingredient for constructing quasi-realistic  superstring
    models in lower dimensions.  As we have seen,  they
  play  also a  crucial role  in  the study of the duality between
    superstring models  and others theories;  in particular in our
    proposition
  given in (2.6).
      A large class of these manifolds are usually  constructed
       as hypersurfaces in toric varieties
        and are nicely described using toric
         geometry technics.    Calabi-Yau
         threefolds are the most important geometries in string
         theory as well as   theirs mirrors
        which   can  been used in   the type II duality,  the geometric
  engineering, 
         F-theory-heterotic duality and this proposed work \cite{31,32,33,34}.   For this reason, we will
  focus our attention
          on  this special   Calabi-Yau geometry.  In particular,
          we develop a tricky  way for getting the mirror
   Calabi-Yau manifolds    using technics of  toric geometry.
   In  this way, we give the mirror   manifolds as hypersurfaces in
   four dimensional weighted  projective  spaces $\bf WP^4$ of weights
   $(w_1,w_2,w_3,w_4,w_5)$. As we will see,
    this
   construction is based on the solving of the  mirror constraint
   equations involved in the toric geometry in terms of the toric
   data of the Calabi-Yau three folds  in  M-theory compactifications. Before
  doing
   this,  let us  start by  describing  $M_3$ in  toric
   geometry framework  and its relation to linear
  sigma model. This will be   useful for our later analysis on
  geometries involved in the  duality between M-theory on $G_2$
  manifolds and F-theory compactifications on elliptically fibred
  Calabi-Yau fourfolds. We first note that toric geometry is a good
  tool for describing the essential one needs about the
  $n$-dimensional Calabi-Yau manifolds and their mirrors involved in
  the previous study.  Roughly speaking, toric manifolds are complex
  $n$-dimensional manifolds
   with $T^n$ fibration over $n$ real dimensional  base spaces with boundary
  [39,40,41,42].
    They exhibit
    toric actions $U(1)^n$  allowing  to encode the
  geometric properties of the complex spaces in terms of simple
  combinatorial data of polytopes ${\Delta}_n$ of the $R^n$ space.
  In this correspondence, fixed points of the toric actions $U(1)^n$
  are associated with the vertices of the polytope ${\Delta}_n$, the
  edges are fixed one dimensional lines of a subgroup $U(1)^{n-1}$
  of the toric action $U(1)^n$ and so on.  Geometrically, this means
  that  the $T^n$ fibres   can degenerate over the boundary of the
  base.  Note that in the case where the base space is compact, the
  resulting toric  manifold  will also be compact.  The beauty of
  the toric representation is that it permits to learn the essential
  about the geometric features of toric manifolds by simply knowing
  the toric data of the corresponding polytope ${\Delta}_n$,
  involving the toric vertices and  the Mori vector weights. \\
  A
  simple example of toric manifold is $ \bf C^n$, which can be
  parameterized by $ z_i= |z_i| e^{i\theta_i}, i=1,\ldots,n$ and
  endowed with  Kahler form given by
  \be
  J= id{\bar z}_i\wedge dz_i= d(|z_i|^2)\wedge d\theta_i. \ee This
  manifold admits  $U(1)^n$  toric actions \be z_i\longrightarrow
  z_i e^{i\theta_i} \ee with fixed locus at $z_i=0$. The geometry of
  $\bf C^n$  can be represented by a  $T^n$ fibration over a
  $n$-dimensional real space parameterized by $|z_i|^2$.  The
  boundary of the base is given by the union of the hyperplanes
  $|z_i|^2=0$.  We have given a very simple example of  toric
  manifolds. However, toric geometry is also a very useful for the
  building ( local) Calabi-Yau manifolds,  providing  a way for
  superstring theory to  interesting physics in  lower dimensions in
  particular four dimensions. An interesting examples are the
  asymptotically local Euclidean ( ALE ) space with $ADE$
  singularities.   These   are   local   complex two dimensions
  toric variety with a $SU(2)$ holonomy.  However, for latter use,
  we will restrict ourselves to
   local three complex Calabi-Yau manifolds $M_{3}$ and their mirrors noted
  $ W_3$ with a $SU(3)$ holonomy.  The toric Calabi-Yau $ M_3$,
  involved in M-theory compactification, can be represented by  the
  following algebraic equation
  \begin{equation}
    \sum \limits _{i=1}^{r+3} Q_i^a |z_i|^2=R^a,
   \end{equation}
   together with the local Calabi-Yau condition
  \begin{equation}
  \sum_{i=1}^{r+3}Q_{i}^{a}=0,\qquad a=1,...,r.
  \end{equation}
  In eqs (3.3-4),   $Q_{i}^{a}$  are integers defining the  weights
  of the toric actions of the complex  manifold in which the $ M_3$
  is embedded. Actually these equations, up some details on $
  Q_i^a$, generalize the one of the weighted  projective space with
  weights   $ Q_i$ corresponding to $r=1$. Each parameter $R^a$ is a
  Kahler deformation  of the Calabi-Yau manifolds. The above
  geometry can be encoded
    in a toric diagram $ \Delta(M_3)$ having $ r+3$  vertices $ v_i$ generating a
   finite  dimensional sublattice of the  $\bf Z^5$ lattice and satisfying the
  following
   $ r$ relations given by
  \begin{equation}
  \sum_{i=1}^{r+3}Q_{i}^{a}{v}_{i}=0,\qquad a=1,...,r,
  \end{equation}
  with  the local Calabi-Yau condition (3.4).\\ Equations (3.3) and
  (3.5) can be related to the so called  D-term potential of two
  dimensional  $ N=2$ sigma model,  for putting the discussion on a
  physical framework \cite {39}. Indeed associating the previous
  variables $z_i$, or $v_i$ vertices  in toric geometry language, to
  $(\phi_i)$ matter fields and interpreting the $ Q_i^a $ integers
  as the quantum charges the $(\phi_i)$'s under a $ U(1)^r$
  symmetry, then the toric Calabi-Yau $M_3$ is now the moduli space
  of  $2D$ $N=2$ supersymmetric linear sigma model. The $ Q_i^a $'s
  obey naturally the neutrality condition, being equivalent to
  $c_1(M_3)=0$,   which means that the theory flows in the infrared
  to a non trivial superconformal model \cite {39,40}. In this way,
  equation (3.3) can be identified with  the  D-flatness conditions
  namely
       \begin{equation}
    \sum \limits _{i=1}^{r+3} Q_i^a |\phi _i|^2=R^a.
   \end{equation}
     In these eqs,  $R^a$  are FT terms which  can be complexified by the theta
  angles  as follows
  \begin{equation}
    t^a=R^a+\theta^a,
   \end{equation}
  where $\theta^a$  have similar role  of the B field in the string
  theory compactification on Calabi-Yau manifolds. The number of the
  independent FI parameters,  or equivalent the number of U(1)
  factors,  equal $h^{1,1}(M_3)$.

  \subsection{ Solving the mirror  constraint eqs for  $W_3$}
  Toric  geometry has been adopted to discussing mirror symmetry as
  well. The latter exchanges the Kahler structure parameters with
  the complex structure parameters. In general, given a toric
  realization of the  manifold $M_3$, one can build its mirror
  manifold $W_3$. This will be  also a toric variety which is
  obtained from $M_3$ by help of mirror symmetry.  In this study, we
  will  use the result of mirror symmetry in sigma model where the
  mirror Calabi-Yau $W_3$ will be  a LG  Calabi-Yau superpotentials,
            depending on the number   of  chiral  multiples and gauge fields
             of dual sigma model on $M_3$.   A tricky way   to   write down the
  equation of
             LG  mirror Calabi-Yau
              superpotential  is to  use
                 dual chiral fields $Y_i$ related to
                  sigma model  fields such that \cite{41,42,43,44}
  \be
   \textrm{Re}\; Y_i= |\phi_i|^2,\quad  i=1,\ldots,k,
  \ee
   and define the   new variables  $y_i$  as  follows $ y_i=e^{-Y_i} $. The  $
  W_3$ LG mirror
   Calabi-Yau   superpotential  takes the form
  \be
   \sum \limits _{i=1}^{r+3} y_i=0,
  \ee
    where the  fields  $y_i$  must satisfy, up absorbing the complex
    Kahler parameters $t^a$, the following constraint equations
  \be
  \prod \limits _{i=1}^{r+3} y_i^{Q^a_i}=1, \quad  a=1,\ldots,r. \ee
  In toric geometry language, this means that the relation between
  the toric vertices of $M_3$  map to relations given by (3.9-10).
  To find an explicit  algebraic equation for the local  mirror
  geometry, one has to solve the constraint equation (3.10). It
  turns out that there many ways to solve these constraint
  equations. Here we
     present a tricky  way, inspired from  the  Batyrev papers   \cite{45,46} and
  \cite{33},
      using the toric geometry representation of sigma
     model of $M_3$. This method can be proceed in some steps. First,
      we note that  the $y_{i}$'s
      are not all independent variables, only $4$ of them do.  The latters can be
  thought as
  local  coordinates  of  the  weighted projective space ${\bf
  WP^{4}}(w_1,w_2,w_3,w_4,w_5)$  which parameterized by the
  following five homogeneous  variables
   \be
   x_\ell=\lambda^{w_\ell}x_\ell,\quad  \lambda\in
  C^*, \ell=1,\ldots,5. \ee   For instance, the  four local
  variables can be obtained from the homogeneous ones using some
  coordinate patch. If
   $x_5=1$,   the other $ x_\ell$ variables behave  as  four
  independent gauge invariants under $C^*$ action of ${\bf
  WP^{4}}(w_1,w_2,w_3,w_4,w_5)$.  The second steep in our program is
  to find relations  between the $y_i$'s and the $x_i$ variables. A
  nice way of obtaining  this  is based  on the using  the  toric
  data of the M-theory Calabi-Yau geometry   for  solving  the
  mirror constraint eqs (3.10). In this method, the mirror geometry
  $W_3$ will be defined as a $D$ degree  homogeneous hypersurfaces
  in ${\bf WP^{4}}$ with the following form
  \begin{equation}
  p_{D}(x_{1},x_{2},x_{3},x_4,x_{5})=0
  \end{equation}
   satisfying \begin{equation}
  p_{D}(\lambda^{w_\ell}x_\ell)= \lambda^D p_{D}(x_\ell).
  \end{equation}
  To write down  the explicit formula of this equation, one has  to
  solve the mirror constraint equation (3.10) in terms  of the
  $WP^4$ toric data. To that purpose, we consider a   solution of
  the dual toric manifold $M_3$ of the form
  \begin{equation}
  \sum_{i=1}^{r+3}Q_{i}^{a}n_{i}^{\ell }=0,\qquad a=1,...,r;\qquad
  \ell =1,...,5,
  \end{equation}
   where $n_{i}^{\ell }$ are integers  specified later on. In patch coordinates
  $x_5=1$,   one
    can parameterize the $y_{i}$
   gauge invariants in
   terms of the $x_i$  as follows
  \begin{eqnarray}
  y_{i}=x_{1}^{(n_{i}^{1}-1)}x_{2}^{(n_{i}^{2}-1)}x_{3}^{(n_{i}^{3}-1)}x_{4}^{(n_{i}^{4}-1)}x_{5}^
  {(n_{i}^{5}-1)}=\prod\limits_{\ell =1}^{5}x_{\ell }^{(n_{i}^{\ell
  }-1)},
  \end{eqnarray}
  where  the deformation given by $y_0=1$  correspond  $
  (n_{0}^{\ell }-1)=0$. Using   the toric  geometry  data of $M_3$,
  (3.10) is trivially satisfied by (3.15).
   Another thing we need in this analysis is that the $y_{i}$
  variables should be thought  of as gauge invariants under the
  ${\bf WP^{4}}(w_1,w_2,w_3,w_4,w_5)$ projective  action  given by
  (3.11). Indeed under this transformation,   the monomials   $y_{i}$
  transform as
  \be
  y_{i}=\Pi _{\ell =1}^{5} x_{\ell }^{(n_{i}^{\ell }-1)} \to {
  y_i}'= y_{i}\lambda ^{\Sigma _{\ell }\left( w _{\ell }(n_{i}^{\ell
  }-1)\right) }\ee and thier   invaraince are 
  constrained by
  \begin{eqnarray}
  \sum_{\ell =1}^{5}w_{\ell } &=&D, \\ \sum_{\ell =1}^{5}w _{\ell
  }n_{i}^{\ell } &=&D.
  \end{eqnarray}
  Equation (3.17) is a strong constraint which will be necessary for
  satisfying the Calabi-Yau condition in the mirror geometry; while
  equation (3.18) shows  that the $n_{i}^{\ell }$ integers involved
  in  ( 3.14-15) can be solved in terms of the partitions $\
  d_{i}^{\ell}$ of the degree $D$ of the homogeneous polynomial
  $p_D(x_{1},...,x_{5})$. Since  $
  \sum_{\ell =1}^{5}d_{i}^{\ell }=D,$  one  can see that   $n_{i}^{\ell }=\frac{%
  d_{i}^{\ell }}{w _{\ell }};$ and take,  for  $i=\ell$,  the
  following property
  \begin{equation}
  n_{i}^{\ell}=\frac{D}{w_{\ell }}, \quad i=1,\ldots, \ell.
  \end{equation}
  In this way the $v_i$ vertices  can be chosen as follows
  \begin{equation}
  v_{i}^{\ell }=n_{i}^{\ell }-e_{0}^{\ell }=\frac{d_{i}^{\ell }}{ w
  _{\ell }}-e_{0}^{\ell },
  \end{equation}
   where $e_{0}^{\ell }=(1,1,1,1,1)$. This shifting will not influence to the
  toric
   realization (3.15)
    due to the Calabi-Yau condition (3.4).    In this way  the first  $6$
   vertices  and the corresponding monomials can be thought of  as follows:
  \begin{eqnarray} v_{0} &=&(0,0,0,0,0) \to \prod_{\ell
  =1}^{5}\left( x_\ell\right)  \nn\\ v_{1} &=&(\frac{D}{w
  _{1}}-1,-1,-1,-1,-1) \to  x_{1 }^{\frac{D}{w _{1 }}} \nn
  \\ v_{2} &=&(-1,\frac{D}{w _{2}}-1,-1,-1, -1) \to x_{2 }^{\frac{D}{w _{2}}}  \\
  v_{3}
  &=&(-1,-1,\frac{D}{w _{3}}-1,-1,-1) \to x_{3 }^{\frac{D}{w _{3 }}}
  \nn\\ v_{4} &=&(-1,-1,-1,\frac{D}{w _{4}}-1,-1) \to x_{4
  }^{\frac{D}{w _{4 }}} \nn\\ v_{5} &=&(-1,-1,-1,-1,\frac{D}{w
  _{5}}-1) \to x_{5}^{\frac{D}{w _{5 }}}.\nn
  \end{eqnarray}
   Using  all the toric vertices in the M-theory geometry, the corresponding
  mirror polynomial
   should involve in the F-theory context takes the  following form
  \begin{equation}
  \sum_{\ell =1}^{5}x_{\ell }^{\frac{D}{w _{\ell }}}+a_0\prod_{\ell
  =1}^{5}\left( x_\ell\right)+  \sum_{i =7}^{r+3}a_i\prod_{\ell
  =1}^{5}x_\ell^{n^\ell_i}=0,
  \end{equation}
   where  the $a_i$'s are complex moduli of the LG Calabi-Yau
   mirror superpotentials. For later use, we take $a_i=0$ and so the above
  geometry
  reduces to
   \be
  \sum_{\ell =1}^{5}x_{\ell}^{\frac{D}{w_{\ell }}}+a_0\prod_{\ell
  =1}^{5}\left( x_\ell\right)= 0.\ee Actually, this  geometry
  extends the quintic  hypersurfaces in  the ordinary $\bf P^4$
  projective space.
  \section{
     $\bf Z_2$  symmetry  in  Toric geometry framework }

   \subsection{  $\bf Z_2$ realization and K3 fibration}
   Here we would like to discuss    the $\bf Z_2$ realization  involved in
    the duality (2.6) using toric geometry tools.  The latter, in the $G_2$
   holonomy sense, will  act  on $M_3$, on   the circle as the
   inversion,  on the $T^2$  and  on  $W_3$ in the F-theory
   compactification. Due to richness of possibilities of the $\bf  Z_2$
   action in the Calabi-Yau manifold, we will focus our attention
   herebelow on giving a  new   $\bf Z_2$  transformation  acting  on the toric  geometry variables. To this  purpose,  we first note that  any
   local
   Calabi-Yau $M_3$,
  described by the  toric   linear sigma model on  (3.3) is, up some
  details,  isomorphic
    to  $C^{r+3}/{C^*}^r $, or equivalently
     \be
     z_i\equiv \lambda^{Q^a_i}z_i,\quad \sum_{i=1}^{r+3}Q_{i}^{a}=0,\quad
  a=1,...,r.
     \ee
     The latter has a $ T^3$  fibration obtained
    by dividing   $ T^{r+3}$  by $U(1)^r$ action generated by a
    simultaneous phase rotation of the coordinates
    \be
    z_i\to e^{iQ^a_i \vartheta^a} z_i,\quad a=1,\ldots,r,
    \ee
   where $\vartheta^a$ are the generators of the $ U(1)$ factors.
   In the Calabi-Yau geometry, plus the reflection on the circle,
   we   will  consider  a   new   $\bf Z_2$ symmetry acting on the toric
  geometry angular
  coordinates.   The latter leads to G2 manifolds with K3 fibration in M-theory side.    To  see this feature,  let  us first consider the simple case  corresponding to  $\bf
     C^3$, that is $r=0$; then we extend this feature to any 3-dimensional  toric varieties.  Indeed, 
  $\bf
     C^3$ has   $U(1)^3$ toric actions
     giving  $ T^3$  fibration in toric geometry realization of  $\bf
     C^3$.  Besides these toric actions,  considering now   a  $\bf Z_2$ symmetry acting    as  follows
\be
  \theta\longrightarrow -\theta_i \qquad i=1,2,3.
   \ee
In this way,  the toric actions  become now 
     \bea
  z_i&\longrightarrow & z_i e^{i\theta_i}\nn\\ \theta_i
  &\longrightarrow &-\theta_i \qquad i=1,2,3,
   \eea
   where $z_i$ are the variables appearing in (3.3). This transformation is quite different  to one given in the leterature [11,12,22],  because it acts on the ongular variables  of complex toric varieties.
   Note, in passing, that one can consider the following   $ \bf Z_2$
   $$
   \theta_i
  \longrightarrow \theta_i +\pi\qquad i=1,2,3;
   $$
   however this transformation is not interesting  from physical argument. Indeed,  first this action has no
   fixed points because  the fixed loci  are naturally identified with brane
  configurations \cite{36}.
   Second, it  does  not leave the following
   constraint equation
    \be
   \theta_1
  + \theta_2 +\theta_3=0,
   \ee
  involved in the determination  of  special Lagrangian manifolds in
  $ \bf C^3$ \cite{43}. \\ Now we return to equation (4.4).  The
  latter   gives   naturally    $ T^3 \o {\bf Z_2}$   as a  fiber space  in the  toric  geometry  realisation of 
    $\bf  C^3$, instead of before  where we have just a  $ T^3$ fibration.
    This feature can be  extended 
      to  any   3-dimensional toric complex manifolds, in particular local   Calabi-Yau threefolds $M_3$. In this way,
     one has   the following
  $\bf Z_2$ symmetry  acting, up the  gauge transformation (4.2), on
  the Calabi-Yau $ M_3$ variables  as follows 
   \bea
  z_i&\longrightarrow & z_i e^{i\theta_i}\nn\\ \theta_i
  &\longrightarrow &-\theta_i \qquad i=1,\ldots,r+3.
   \eea
  Geometricaly,  this transformation   gives   
   a  local Calabi-Yau threefolds with  $T^3\o \bf Z_2$
        fibration. In this  case, plus the toric geometry action fixed points we  have now extra ones coming   from  the orbifold toric  fibration.
  These fixed loci  may  be  identified with brane configurations
  using the interplay between  toric geometry and type II brane
  configurations \cite{36}. For the moment we ignore the brane
  description and return to the geometric interpretation of the
  (4.6).  Indeed, it  is easy to see that, together with the action
  on the circle,
      these  new toric actions  lead  to  G2 manifolds  with  $T^4\o{\bf
      Z_2}$ fibration,  being as the  orbifold limit of
  the $K3$ surfaces, in  the  M-theory  compactifications. The geometry given by (2.3)  can be now  viewed as a  $G_2$ manifold with
       $K3$ fibration. In this way, we are able to find  heterotic superstring  dual  models which could be used  to support our  proposed   duality  (2.6).   Indeed, the  moduli space of  smooth
      compactifications    can be
      obtained from the  one  of $K3$ followed by an extra compactification  on a  three  dimensional space $Q_3$,  down to four dimensions \footnote{ The
      $G_2$ holonomy  and the Calabi-Yau condition of $K3$ require
      that $b_1(Q_3)=0$.}. This can be related directly  to
    the   heterotic superstring by fibering  the M-theory/heterotic
      duality in seven dimensions on the same base $Q_3$.  Locally, the moduli
      space of this compactification  should have the following  form \be
       {\cal M}(K3) \t  {\cal M}(Q_3)
       \ee
       where ${\cal M}(K3)$ is the  moduli space of  the M-theory  on
       $K3$ in seven dimensions $$
   {\cal M} (K3)= {\bf R}^+ \t{SO(3,19)\o {SO(3)\t SO(19)}}
      $$  which is exactly  the moduli space of heterotic strings  on $ T^3$. 
     ${\cal M}(Q_3)$  describes the physical   moduli  coming after  the extra
     compactification on
     $X_3$.  This compactification describes  the strong limit of   heterotic superstrings  on Calabi-Yau threefolds $Z$,  being a   $ T^3$ fibration on $Q_3$.    More recently,   it was shown that   M-theory on  a   $G_2$ manifolds  with K3 fibration  can give  much more interesting  physics that other  superstrings  derived models;  in particular  it  leads to a theory  similar to  a four dimensional grand unified model [20]. 
Alternatively,     the  $ T^4 
\over {\bf  Z_2}$  fiber space   is locally  isomorphic to  $ {\bf C^2}\o  {\bf Z}_2$  known by  $A_1$ singularity,   which    can be
  determined algebraically in terms of the ${\bf Z}_2$ invariant
  coordinates on ${\bf C^2}$ as follows
  \be
     z^2=xy.
  \ee M-theory on this local geometry singularity corresponds to two
  units of D6 branes.
     For general case where we have $A_n$ singularity, this geometry
     is equivalent to $n+1$ D6 branes of type IIA superstring. In this way, the above compactification may have a  four dimensional  interpretation in terms of type IIA  D 6 branes.\\  Before going  to F-theory, Now we want to discuss the Rahm
  cohomology of the M-theory quotient
      space. The  Calabi-Yau condition
       $b_1(M_3)=0$ and the $G_2$ holonomy condition  require  that $b_1(X_7)=0$.  Using eqs (4,6), the Kahler form now is  odd under the
      above $\bf Z_2$ symmetry. Since there are no invariant 2 forms, we have the
  following
      constraint for  the quotient space
      \be
      b_2=0
      \ee
      However there are some  invariant three forms; one  type of them    is  
  given by
      \be
      \phi= J\wedge dx
      \ee
      where  $J$ is the Kahler form  on $ M_3$ and  $x$ is a   real coordinate
  parameterizing
      the circle.  The number of these  forms  is given by $h^{1,1}(M_3)$ being
      the dimension of  the
       complexified Kahler moduli space of $M_3$.\\
  Now we go on to the F-theory  to give the corresponding
  geometries. Instead of being general,  we will consider
    a concrete example  describing   the mirror quintic hypersurfaces obtained by
  taking
    \be
    w_\ell=1\qquad \forall  \ell.
    \ee
   In this way,   the general mirror geometry (3.23) reduces to
  \be
  \sum_{\ell =1}^{5}x_{\ell}^{5}+a_0\prod_{\ell =1}^{5}\left(
  x_\ell\right)= 0.\ee
     This equation   has    any  direct toric description of
    the  $\bf P^4$  in which it is embedded.  However a toric
    realization may be recovered if we consider the limit
    $a
  _0\to \infty$ in the mirror description. In this limit,  the defining equation of
  the
   mirror geometry  becomes approximately, up scaling out the $a_0$,
   as
    \be
    x_1\ldots x_5=0.
    \ee
    This equation   can be solved by taking one or more  $ x_i=0$.
    In toric geometry language, this solution
    describes the union of the boundary faces of 4-simplex defining  the polytope
  of the
    $\bf P^4$ projective space \cite {36}.  In this case,   the  F-theory mirror
  quintic  is   a $T^3$
    fibration over 3-dimesional real space defined by   the boundary faces of
  4-simplex
    of  $\bf P^4$ . It consists of the intersection of 5 $\bf P^3$'s
     along 10 $\bf P^2$'s. This geometry has now a $ U(1)^3$ toric action which
      can be deduced from  the
     ones  of $\bf P^4$  and can be thought of   as follows
     \bea
    x_i&\to & e^{i\theta_i } x_i,\quad i=1,2,3\nn\\
     x_i&\to  & x_i \quad i=4,5
    \eea
    In the  F-theory geometry,
  $\bf Z_2$ symmetry   will act on  the mirror  Calabi-Yau
  threefolds as follows
   \bea
  x_i&\longrightarrow & x_i e^{i\theta_i}, \quad i=1,2,3\nn\\
  \theta_i &\longrightarrow &-\theta_i.
   \eea
  Here we repeat the same analysis of M-theory. In this case,
       the F-theory  geometry  can
    have also  $ {T^4\o
     \bf Z_2}$  fibration over  a four dimensional base space.  However,   a naive
  way to
      get an   elliptically  K3 fibration, being the relevant geometry in the
  F-theory
       compatification
    for  obtaining  $N=1$ in four dimensions, is  to suppose that the $\bf Z_2$
  symmetry
    acts trivially on the torus of $ W_4$  Calabi-Yau fourfolds. In this way,  the 
  $ {T^4\o
     \bf Z_2}$  fibration reduces  to $ T^2\times { T^2\o
     \bf Z_2}$ which is an elliptic model  in the context of  F-theory compactifications. 
  This compactification
     should be interpreted in terms of type IIB on  $ { T^2\o
     \bf Z_2}$.    By this  limit,   one can see that the   orbifold
     (2.4)
      gives an elliptic  K3 fibration in F-theory  compactifications.
    In  M-theory context, this  geometry can be obtained using
       the following
   factorization in the Narain lattice
     \be
     \Gamma^{19,3}= \Gamma^{18,2}+ \Gamma^{1,1}
     \ee
     having a  nice interpretation in terms of
  the  action of  $\bf Z_2$ symmetry  on the moduli space of K3
     surfaces \cite{47}.

     \subsection{ More on the  $\bf Z_2$ action }
  We would like  to give comments regarding   a particular
  realization of $\bf Z_2$  symmetry  when it
    acts trivially on the
    Calabi-Yau threefolds. In this way, the
    geometries (2.3) and (2.4) reduce  respectively to
    \bea
  X_7={S^1 \over {\bf Z_2}}\times M_3\\ W_4= {T^2 \over {\bf
  Z_2}}\times W_3.
    \eea
    In M-theory,  this compactification may be thought of  as    the
  Ho$\check{r}$ava$-$Witten
    compatification  on spaces
  of the form ${S^1 \over {\bf Z_2}}\times Y$, where $Y=M_3$ is a
  Calabi-Yau threefolds \cite{48}. M-theory on  this type of
  manifolds gives  rise to $N=1$ supersymmetry in four dimensions,
  having  a weak coupling limit given by the heterotic superstring
  compactified  on $M_3$. Using the toric description  of  $M_3$
  where one has a  $ T^3$ fibration over three dimensional base
  space, the compactification  (4.12) may be related to M-theory on
  $G_2$ manifolds with $K3$ fibration. This  may  be checked by the
  seven dimensional  duality between M-theory on K3 and the
  heterotic superstring on $T^3$.   In the end of this section  we
  want to discuss the F-theory duals of this
   kind of the   compactification. Instead of being general, we will consider
  concrete examples corresponding to $M_3$,  described by $N=2$
  linear sigma model on the canonical
   line bundle over two complex  dimensional toric  varieties.    In F-theory, the mirror map of these geometries are 
  given  by non
   compact Calabi-Yau threefolds LG superpotentials with an equation of the form
  \be
   W_3: \quad f(x_1,x_2)=uv, \ee
   where  $  x_1,x_2$  are  $ C^*$
    coordinates  and  $u,v$ are $ C^2$ coordinates. For illustrating applications,  
  let us give two   examples.\\
     {\bf  (i) $\bf P^2$ projective space}\\
  The first example is  the sigma model on
     the  canonical line bundle over $\bf P^2$. In this way,  the Calabi-Yau
  geometry in  M-theory
     side  is  described by    a   $U(1)$   linear sigma model  with four   matter
      fields  $\phi_i$   with the following   vector charge
  \be
  Q_i=(1,1,1,-3), \ee
   satisfying the Calabi-Yau  condition (3.4). After
  solving the mirror constraint equations (3.10), the corresponding
  $W_3$  LG Calabi-Yau superpotential  in F-theory compactification
  is given by the following equation  \be \quad W_3:\quad
  f(x_1,x_2)= 1+x_1+ x_2+{e^{-t}\o
   x_1x_2}=uv.    \ee
  {\bf (ii)  Hirzebruch surfaces
      $\bf F_n$ }\\
      As a second example, we consider a  local model given by the canonical line
  bundle over the  Hirzebruch surfaces
      $\bf F_n$. Recall  by the way that
        the $\bf F_n$ surfaces are two dimensional toric manifolds,  defined by a
  non-trivial
        fibration of a $\bf P^1$  fiber
            on a $\bf P^1$ base.   The latter  are realized as the vacuum manifold
  of the
           $ U(1) \times U(1)$ gauge theory with four chiral fields with charges
           \bea
            Q_i^{(1)}&=&(1,1,0,-n)\nn\\
           Q_i^{(2)}&=&(0,0,1,1).\eea
  The canonical line bundle over these surfaces is a local
  Calabi-Yau  threefolds  described  by  an
       $U(1)\t U(1)$   linear sigma model  with  five  matter  fields  $\phi_i$  
  with
       two  vector charges \bea Q_i^{(1)}&=&(1,1,0,-n,n-2)\nn\\
  Q_i^{(2)}&=&(0,0,1,1,-2), \eea   These satisfy naturally  the
  Calabi-Yau condition (3.4). After solving (3.10),  the defining
  equation
            for  the   LG  mirror  superpotential becomes
  \be W_3\quad f_n(x_1,x_2)=uv,\ee where $f_n(x_1,x_2)=
  1+x_1+{e^{-t_1} x_2^n\o x_1}+x_2+{e^{-t_2}\o
   x_4}$ and where $t_i$ are complex parameters.\\
  In F-theory side,  eqs (4.21) and (4.24)
    describe   elliptic fibration solutions of (4.19), where the elliptic
    curve
    fiber is given by
    \be  f(x_1,x_2) =0.
    \ee
  By introducing an extra variable,  this elliptic curve can have a
  homogeneous representation described by the  cubic polynomial in
  ${\bf P^2}$,  with the   general
   form as follows
  \be
  \sum\limits_{i+j+k=3}a_{ijk}x^iy^jz^k=0,  \ee which  can be
  related,  up some limits in the complex structures, to  the
   following Weierstrass form
  \be
  y^2z=x^3+axz^2+bz^3. \ee This form plays an important role
     in the construction of elliptic Calabi-Yau  manifolds  involved in
     F-theroy compactifications \cite{21,22,44}. \\
     The elliptic K3 fibration of (4.8) may be
   obtained by using only the orbifold  of  the torus in F-theory.
   A  tricky way
    to see this is  as follows. First, we  consider the elliptic curve  $T^2$ as a
      fiber circle $ S^1_f$,  of radius $R_f$,   fibered on a  $
   S^1_b$ base circle,  of radius $R_b$. After that  we take  a   $\bf Z_2$
    symmetry  acting
   only on the  $S^1_b$ and leaving the  $S^1_f$ fiber invariant. In this way, the
      orbifold
    $ T^2\o {\bf Z_2}$ can be
    seen as
   a  $ S^1_f$ bundle over   a line segment, with
   two fixed points zero and $\pi R_b$.  This space, up some
   details, has similar feature of the toric realization of  $ \bf P^1$.
    To see this, assuming that   the  radius of the $ S^1_f$ fiber  vary on the
   base  space  as
   follows
   \be
   R_f\sim \sin {x \over R_b}
   \ee
  where $x$ is the coordinate on the interval which runs from zero
  to $\pi R_b$. In this case,  the $ S^1_f$ can  shrink
   at the two end-points of the line segment. With this assumption, the resulting
   geometry is
  identified with the toric  geometry realization of the $\bf P^1$
   [39,40]. By this limit in the $\bf Z_2$   orbifold,
  the F-theory geometry given by (4.18) can be viewed  as  a $\bf
  P^1$ fibration over  on $ W_3$. Since $W_3$ is elliptic model, 
    this fibration,  in presence of   $
   \bf P^1$,   can give an elliptic $ K3$ fibration  which is the relevant
  geometry
    for  getting
   $N=1$  models in four dimensions from F-theory compactifications.
   \section{ Discussions and conclusion }
   In this paper, we have studied,  geometrically,  the $N=1$ four dimensional
   duality between M-theory on $G_2$ manifolds  and F-theory on
   elliptically fibred Calabi-Yau fourfolds with $SU(4)$ holonomy.
   In this study, we have considered M-theory on spaces of the form $ {S^1 \times
   M_3}\over {\bf Z_2}$, where $M_3$ is  a Calabi-Yau threefolds
   described physically by $ N=2$ sigma model in two dimensions. In
   particular, using  mirror symmetry, we have discussed the
   possible duality  between M-theory on such spaces and  F-theory
   on $ {T^2 \times
   W_3}\over {\bf Z_2}$, where $ (M_3, W_3)$ are mirror pairs in type
   II superstring compatifications on Calabi-Yau threefolds.
     In this  work, our results are summarized as
   follows:\\
   (1) First,  we have developed   a way for getting LG  Calabi-Yau
   threefolds $W_3$ mirror to sigma model on toric Calabi-Yau $
   M_3$. This method is based on solving the mirror constraint eqs
   for LG theories in terms of the toric data of sigma model on $
   M_3$. Actually, this way gives directly the right dimensions of
   the mirror geometry.  In particular,  we have shown that the mirror LG
   Calabi-Yau threefolds $W_3$
   can be described as hypersurfaces in four dimensional weighted projective
  spaces
  $\bf WP^4$, depending  on the toric data of $M_3$.  In this way,
  the $\bf WP^4$ may be determined in terms of the toric geometry
  data of $M_3$.\\
   (2) Using these
  results, we have shown, at least, there exist
     two classes of F-theory duals of M-theory on $ {S^1 \times
   M_3}\over {\bf Z_2}$. These classes depend on the possible
   realizations of the ${\bf Z_2}$ actions on $ M_3$. In the case,
   where ${\bf Z_2}$ acts no trivially on $ M_3$, we have given  a toric
   description of  the above mentioned duality. In particular, we
   have proposed a special ${\bf Z_2}$  symmetry acting on the toric
   geometry angular variables. This way gives seven manifolds with K3
   fibrations of $G_2$ holonomy.  For the case, where ${\bf Z_2}$ acts trivially
  on $
   M_3$, we have given some illustrating examples of the F-theory duals of
   M-theory.
   Finally,   since every supersymmetric intersecting brane dynamics is expected
   to lift to M-theory on local $ G_2$ manifolds \cite{49},
   it would be interesting to explore this physics    using intersecting D 6-7
  branes in
    Calabi-Yau
    manifolds. Moreover, the  $N=1$ model studied in this work  can be related to
  heterotic superstrings on Calabi-Yau threefolds with the  elliptic
  fibration  over del Pezzo surfaces $ dP_k$, the $\bf P^2$'s blown up in
  $k$ points with $k\leq 9$. In a special limit these surfaces may
  be specified by the elliptic fibration $ {W_3\o {\bf Z_2}} \to
  dP_k$ with $\bf P^1$ fibers.  It should be also interesting to explore
  physical realizations,  via branes, of this fibration. Progress in
  this direction will be reported elsewhere.
   \\
   \\
   {\bf
  Acknowledgments}\\
   I would like  to thank many peoples. I would like  to  thank
    Instituto de Fisica Teorica, Universidad Autonoma de Madrid for kind
  hospitality
     during the preparation
     of this work.  I am  very grateful to  C. G\'omez   for discussions,
     encouragement and   scientific helps.
     I would like to thank  A.M. Urenga for  many valuable discussions and
  comments on this work  during my  stay at
     IFT-UAM.
     I am  very grateful to E. H. Saidi  for discussions, encouragement and 
  scientific helps.
      I would like to thank the  organizers of the Introductory School on String
  theory (2002),
      the Abdus
  Salam International Centre for Theoretical Physics, Trieste,
  Italy, for kind hospitality. I would  like to thank  K. S. Narain,
  M. Blau  and T. Sarkar
    for discussions at ICTP.  I am  very grateful to  V. Bouchard  for the
  reference [14].
     I would like to thank J. McKay and A. Sebbar
      for discussions, encouragement and  scientific helps. This
    work is  partially  supported by SARS, programme de
  soutien \`a la recherche scientifique; Universit\'e Mohammed
  V-Agdal, Rabat.  I would like to thank my family  for helps.

   \end{document}